\documentstyle[11pt,a4wide]{article}

\begin{document}
\begin{flushright}
ASITP-94-41
\end{flushright}
\vspace{.5cm}
\centerline{\bf\large{\bf{Darboux Transformations for Supersymmetric  Korteweg
- de Vries Equations }}}
\vspace{1in}
\centerline{\bf\large Q.\ P. \ Liu}
\centerline{\bf\large Institute of Theoretical Physics, Academia Sinica,}
\centerline{\bf\large P.\ O.\ Box 2735, Beijing 100080, China}
\newcommand{\be}{\begin{equation}}
\newcommand{\ee}{\end{equation}}
\newcommand{\ba}{\begin{array}}
\newcommand{\ea}{\end{array}}
\def\p{\partial}
\def\alf{\alpha}
\def\bi{\beta}
\def\es{\epsilon}
\def\la{\lambda}
\vspace{0.5in}\begin{center}
\begin{minipage}{5in}
{\bf ABSTRACT}\hspace{.2in}We consider the Darboux type transformations for the
spectral problems of supersymmetric KdV systems. The supersymmetric analogies
of Darboux and Darboux-Levi transformations are established for the spectral
problems of Manin-Radul-Mathieu sKdV and Manin-Radul sKdV. Several B\"acklund
transformations are derived for the MRM sKdV and MR sKdV systems.\par
\end{minipage}
\end{center}
\vspace{.5in}
\vfill\eject
\par
\section{Introduction}
Soliton systems, such as KdV equation, Sine-Gordon equation and Nonlinear
Schr\"{o}dinger equation, have been found to enjoy many remarkable properties,
for example, they may be solvable via Inverse Scattering Transformation, have
Bi-Hamiltonian structure, have infinite number of conservation laws(symmetries)
and B\"{a}cklund transformations(BT), etc.. Among many approaches to construct
BT for a given system, so-called Classical Darboux Transformation(CDT) is the
most direct and yet elementary one. In words, CDT establishes a relationship
between solutions of a couple of one dimensional Schr\"{o}dinger equations or
recover some covariance of it[14].\par
CDT is first used by Wadati, Sanuki and Konno[20] to obtain soliton solutions
for the KdV equation. We also notice that the interesting work of Levi,
Ragnisco and Sym[10], which shows that CDT is equivalent to the important
dressing transformations.\par
Apart from DT, a new version of Darboux transformation exists. Indeed, Levi[9]
introduced a transformation, now known as Darboux-Levi's
Transformation(DLT)[1].  Most recent works along this line include DT for KP
hierarchy[19][2] and DT in higher dimensional case[19][6]. We refer to Matveev
and Salle's book[15] for more examples and applications of DT(see also the
review paper in ref.[6]).\par
 The present paper is intended to generalize DT to the context of
supersymmetric case. The super KdV equations are introduced recently. We
mention the Kupershmidt's version[8] and Manin-Radul's version[11]. While the
former is just a fermionic extension of the KdV system, the latter is a genuine
supersymmtric KdV(sKdV)[12]. Here, we are only concerned with the
supersymmetric case. It is interesting to notice that various structures of
sKdV systems are exploited. We just cite the Bi-Hamiltonian theory[18][4][7],
the relationship between the second Poisson brackets and $N=1$ superconformal
algebra[13], Bilinear form[16] and sKdV can be embedded in the supersymmetric
Yang-Mills[5]. We will show that CDT and DLT have their analogies in
supersymmetric context. We will see that DT and DLT here is less general than
the ordinary case due to the pressure of non-commutativity, but they do have
interesting consequence: we are able to derive BT for sKdV equations.\par
The paper is arranged as follows. Next section includes two theorems on DT for
MRM sKdV and MR sKdV. The following section is intended to the super version of
DLT. In section 4, we calculate some exact solutions for both MRM sKdV equation
and MR sKdV equation by means of the Darboux transformations we established in
section 2 and last section is a brief summary and some comments. \par
{\em Connection}: we follow the usual convection: denoting the fermionic(odd)
filed variables by Latin letters and bosonic(even) field variables by Greek
letters. Also, we always use Greek letters to denote the wave functions whose
parities will be stated clearly if necessary.\par
\section{Darboux Transformations}
Let us start with the Manin-Radul-Mathieu sKdV system[11,12]:
\be
\alf_t=\frac{1}{4}(\alf_{xxx}+3(\alf D\alf)_x).
\ee
Where $D=\theta \p +\p_{\theta}$.
The Lax operator for (1) is:
\be
{\bf L}_{mrm}=\p^2+\alf D,
\ee
it is easy to see that the equation(1) is equivalent to:
\be
{\bf L}_{mrm_{t}}=[{\bf P}_{mrm}, {\bf L}_{mrm}],
\ee
where ${\bf P}_{mrm}=\p^3+\frac{3}{2}\alf D^3+\frac{3}{4}\alf_x D$.

For the system(1), we have:
\par
{\em Proposition 2.1}.\par
If $\phi$ satisfys the equation:
\be
\phi_{xx}+\alf(D\phi)=\la \phi,
\ee
and $ \Lambda$ is a solution of the equation(4) with zero energy. Taking
\be
{\hat{\phi}} =(\p-\frac{\Lambda_{x}}{D\Lambda}D)\phi,
\ee
then ${\hat{\phi}}$ satisfys:
\be
{\hat{\phi}}_{xx}+{\hat{\alf}}(D{\hat{\phi}})=\la {\hat{\phi}},
\ee
with
\be
{\hat{\alf}}-\alf=2(\frac{\Lambda_{x}}{D\Lambda})_x.
\ee
\bigskip
Proof. A straightforward calculation.\par
\bigskip
{\em Remarks}:\par
(i). Here, that $\Lambda$ is a solution of the equation (4) with zero energy
is  sufficient  but not necessary. In fact, if we take $\Lambda$ as a solution
of the equation $\phi_{xx}+\alf(D\phi)=\la_0 \phi$, $\la_0 \neq 0$, then this
requires $\Lambda \Lambda_x=0$.\par

(ii). The proposition makes sense only if $\Lambda $ is a odd(fermionic)
function.\par
\bigskip
Next, we study  application of the Proposition 2.1. Actually, we are able to
derive a B\"{a}cklund transformation for MRM sKdV. The equation (7) gives us:
\be
\delta =\frac{1}{2}\p^{-1}({\hat{\alf}}-\alf)+\eta,
\ee
where we used the notation $\delta=\frac{\Lambda_{x}}{D\Lambda}$ and $\eta$ is
a odd constant.\par
Also, using the equation $\Lambda_{xx}+\alf (D\Lambda)=0$, we find:
\be
\delta_x +\delta (D\delta) +\alf=0,
\ee
and substitution of (8) into (4) leads to:
\be
{\hat{\alf}}+\alf +\frac{1}{2}\alf
(\p^{-1}({\hat{\alf}}-\alf)+2\eta)D^{-1}({\hat{\alf}}-\alf)=0.
\ee
We notice that by redefining the variable, we may put the BT(10) in the
differential form:
\par
\begin{center}
$ {\hat{\bi}}_x+\bi_x+ \frac{1}{2}\bi_x[{\hat{\bi}} -\bi
+2\eta][D{\hat{\bi}}-D\bi]=0,$
with $\alf=\bi_x and {\hat{\alf}}={\hat{bi}}_x$.
\end{center}
\par
After the presentation of the Darboux transformation for MRM system(1), we turn
to the analogous results for MR's sKdV equation:
\be
\begin{tabular}{ll}
&$\alf_t=\frac{1}{4}(\alf_{xxx}+3(\alf D\alf)_x+6(\alf u)_x),$\\[2mm]
&$u_t=\frac{1}{4}(u_{xxx}+6uu_x+3\alf_x (Du)+3\alf (Du_x)).$
\end{tabular}
\ee
The corresponding Lax operator is given by:
\be
{\bf L}_{mr}= \p^2 +\alf D +u,
\ee
and the Lax representation is:
\be
{\bf L}_{mr_{t}}=[{\bf P}_{mr}, {\bf L}_{mr}],
\ee
where ${\bf P}_{mr}=\p^3+\frac{3}{2}\alf
D^3+\frac{3}{2}u\p+\frac{3}{4}\alf_xD+\frac{3}{4}u_x.$

As above, we have:\par
\bigskip
{\em Proposition 2.2}.\par
Let  $(\p^2 +\alf D +u)\psi=\la \psi$ and $\Lambda$ is a particular solution of
it with $\la=\la_0$. Taking
\be
{\hat{\psi}}=(D+\delta)\psi,
\quad \delta=-\frac{D\Lambda}{\Lambda},
\ee
then the following equations are satisfied:
\be
{\hat{\psi}}_{xx}+{\hat{\alf}}(D{\hat{\psi}})+{\hat{u}}{\hat{\psi}}=\la
{\hat{\psi}},
\ee
where
\be
{\hat{\alf}}+\alf=-2\delta_x, \quad {\hat{u}}-u=(D\alf)+2\delta\alf+2\delta
\delta_x,
\ee
\bigskip
Proof: A direct calculation.\par
\bigskip
{\em Remarks}:\par
(i). $\Lambda$ is a bosonic(even) function.\par
(ii). Since the operator $(D+\delta)$ is a fermionic (odd) operator, the wave
functions $\psi$ and ${\hat{\psi}}$ have different parities.\par
\bigskip
A B\"{a}cklund transformation may be obtained for MR's sKdV(11) by means of the
proposition 2.2. Let us derive it next.\par
It is easy to see that the $\delta $ can be represented as:
\be
\delta=-\frac{1}{2} \p^{-1} ({\hat{\alf}}+\alf)+\eta,
\ee
where $\eta$ is an integration constant which is fermionic. Using relation
$D\Lambda=-\delta \Lambda$ and the equation $\Lambda_{xx}+\alf
(D\Lambda)+u\Lambda=\lambda_0\Lambda$, we see that $\delta $ satisfys the
following equation:
\be
-(D\delta_x)+(D\delta)^2-\alf \delta +u=\la_0,
\ee
now, substituting the expression(17) into the equation(17) and the second
equation of (16) yields:
\be
\begin{tabular}{ll}
&$\frac{1}{2}((D{\hat{\alf}}+(D\alf))+\frac{1}{4}[D^{-1}({\hat{\alf}}+\alf)]^2+\frac{1}{2}\alf\p^{-1}({\hat{\alf}}+\alf)-\alf \eta +u=\la_0,$\\[2mm]
&$u-{\hat{u}}=(D\alf)+[-\p^{-1}({\hat{\alf}}+\alf)+2\eta]\alf-[-\frac{1}{2}\p^{-1}({\hat{\alf}}+\alf)+\eta]({\hat{\alf}}+\alf),$
\end{tabular}
\ee
which is a B\"acklund transformation for MR's sKdV(11).
\par
As before, we may introduce a new variable and rewrite the above BT in local
form.\par
\section{New Darboux Transformations}
This section is intended to construction of new types of Darboux
transformations for both MRM sKdV and MR sKdV. That is, we will show that
Levi's generalization of Darboux transformation also has its analogies in the
super case.\par
Indeed, for the MRM sKdV case, we obtain:\par
\bigskip
{\em Proposition 3.1}\par
If $\phi$ is a solution of the equation $\phi_{xx}+\alf (D\phi )=\la \phi$ and
$\theta$ is a particular solution of this equation with zero energy. Letting
\be
{\hat{\phi}}=\frac{D^{-1} \Omega}{\theta}, \quad
\Omega=(D\phi)\theta-(D\theta)\phi,
\ee
then, ${\hat{\phi}}$ is the solution of the  equation:
\be
{\hat{\phi}}_{xx}+{\hat{\alf}}{\hat{\phi}}=\la {\hat{\phi}},
\ee
with:
\be
{\hat{\alf}}=\alf -4(\frac{D\theta}{\theta})_x.
\ee
\bigskip
Proof. The calculation involved here is rather tedious. Since the idea employed
is similar to the one used next Proposition, we omit the proof here.\par
\par
\bigskip
{\em Remarks}.\par
(i). The particular solution $\theta$ must be a bosonic function.\par
(ii). This time, the wave function $\phi$ and ${\hat{\phi}}$ have the same
parity.\par
(iii). To be accurate, we must choose the special solution $\theta$ such that
the quantity:\par
$ I=\lambda ({\hat{\phi}}-\phi)\theta
-2\theta_x\phi_x+2\theta^{-1}\theta^{2}_{x}\phi-2(D\phi_x)(D\theta)+2(D\phi)(D\theta_x)+4\theta^{-1}(D\theta_x)\phi(D\theta).$ \par
\bigskip
It is easy to see that the BT in this case reads
\be
(D{\hat{\alf}})-(D\alf
)+\frac{1}{4}[D^{-1}({\hat{\alf}}-\alf)]+\alf[\p^{-1}({\hat{\alf}}-\alf)+4\eta]=0.
\ee
\bigskip
As for the MR sKdV system(11), we get:

{\em Proposition 3.2}
\par
Let $\psi$ and $\theta$ be solutions of the equation $(\p^2+\alf
D+u)\psi=\la\psi$ with the spectral parameters $\la$ and $\mu$ respectively.
Taking
\be
{\hat{\psi}}=\frac{\theta}{\Theta}(D^{-1}(\frac{\Theta
\Omega}{\theta^2}))+k\theta \Theta^{-1},
\ee
with $\Theta$ satisfys the following linear equation:
\be
\Theta_{xx}-2\frac{\theta_x}{\theta}\Theta_x
+(\alf-2\frac{D\theta_x}{\theta}+2\frac{(D\theta)\theta_x}{\theta^2})(D\Theta)=0,
\ee
Then, ${\hat{\psi}}$ is the solution of the equation $(\p^2+{\hat{\alf}}
D+{\hat{u}})\psi=\la\psi$ but with:
\be
{\hat{\alf}}=\alf -2\Gamma_{x},
\quad \hat{u}= u-4(D\gamma)(D\Gamma)-2(D\Gamma)^2+2\gamma
\Gamma_x+2\Gamma\gamma_x+2\Gamma\Gamma_x,
\ee
where $\gamma=-\frac{D\theta}{\theta}$, and
$\Gamma=\frac{D\Theta}{\Theta}$.\par
\bigskip
Proof. Let us calculate the quantity:\par
\be
{\hat{\psi}}_{xx}+{\hat{\alf}}(D{\hat{\psi}})+{\hat{u}}{\hat{\psi}}=
(\mu\theta \Theta^{-1})D^{-1}(\Theta
\theta^{-2}\Omega)+k\mu\theta\Theta^{-1}+I,
\ee
where
\be
\begin{tabular}{ll}
&$I= 2(\theta\Theta^{-1})_x D(\Theta \theta^{-2}\Omega)+\theta
\Theta^{-1}D\p(\Theta\theta^{-2}\Omega)+\alf\theta^{-1}\Omega$\\[2mm]
&$ \quad \quad
-2\Theta^{-1}\theta^{-1}(D\Theta_x)\Omega+2\Theta^{-2}(D\Theta)\Theta_x\theta^{-1}\Omega,$
\end{tabular}
\ee
A long calculation shows that
\be
\begin{tabular}{ll}
&$I=-\Theta^{-1}\Theta_x\psi_x
+\theta^{-1}\Theta^{-1}\Theta_x\theta_x\psi-\Theta^{-1}(D\Theta_x)(D\psi)+\Theta^{-1}(D\Theta)(D\psi_x)+$\\[2mm]
&$\quad \quad
\theta^{-1}\Theta^{-1}(D\Theta)(D\psi)\theta_x-\theta^{-1}\Theta^{-1}(D\Theta)\psi_x(D\theta)-\theta^{-1}\Theta^{-1}\psi (D\Theta)(D\theta)_x+$\\[2mm]
&$\quad \quad
(\psi_{xx}+\alf(D\psi))-\theta^{-1}\theta_{xx}\psi-\alf\theta^{-1}(D\theta)\psi
+\theta^{-1}\Theta^{-1}(D\Theta)_x\psi(D\theta).$
\end{tabular}
\ee
Thus,
\be
{\hat{\psi}}_{xx}+{\hat{\alf}}(D{\hat{\psi}})+{\hat{u}}{\hat{\psi}}=\la
{\hat{\psi}}+\theta \Theta^{-1}[-\la\theta^{-1}\Theta{\hat{\psi}}+\mu
D^{-1}(\Theta\theta^{-2}\Omega)+k\mu+\theta^{-1}\Theta I],
\ee
Now we have to prove that the quantity in the square bracket is a constant. We
denote it as ${\hat{I}}$.
Then tedious calculation shows:
\be
D{\hat{I}}=0,
\ee
thus, ${\hat{I}}$ is a constant. This constant can be made as zero by suitable
choose of integration constant and suitable boundary conditions.
\par
\bigskip
{\em Remarks}.\par
(i). The wave functions $\psi$ and ${\hat{\psi}}$ have the same parity.\par
(ii). A B\"{a}cklund transformation may be derived for the MR sKdV system.
However, it is in a very complicated form.\par

\section{Exact Solutions of Supersymmetric KdV Systems}
Among many applications of Darboux transformations, the most important and
direct one is to construct exact solutions for the related nonlinear
equations[14]. In the following, we will show that some interesting solutions
of sKdV systems may be obtained by means of the propositions 2.1 and 2.2.

In order to use the proposition 2.1 and proposition 2.2 for the sKdV systems,
we have to verify that the time evolution of wave functions are covariant under
the Darboux transformations, i.e., the equations $\phi_t={\bf P}_{mr} \phi$ and
$\psi_t={\bf P}_{mrm} \psi$ are covariant under the Darboux transformations
given in proposition 2.1 and proposition 2.2 respectively. This indeed can be
proved by a long but straightforward calculation.

Let us first consider the MRM sKdV case. The simplest solution of the equation
() is the trivial one: $\alf=0$. With this solution, we find that
\be
\Lambda=\theta (ax+b),
\ee
where a and b are arbitrary(bosonic) constants. Thus, the solution in this case
is:
\be
\alf=2(\frac{\theta a}{ax+b})_x,
\ee
this is a stationary solution of MRM sKdV equation(1).

For the MR sKdV system(11), we also start with the trivial solution $\alf=u=0$.
Then, choosing $\Lambda$ in the form:
\be
\Lambda=(1+\theta \xi)cosh(kx+k^3t),
\ee
where $\xi $ is a fermionic constant and $\la=k^2$. The solution of () in this
case reads:
\be
\alf=2\frac{k\theta}{(1+\theta\xi)cosh^2(kx+k^3t)},\quad
u=2\frac{k\xi\theta}{(1+\theta\xi)^{2}cosh^2(kx+k^3t)}.
\ee
Interestingly if we let $\theta=\xi$, we obtain a solution for MRM sKdV
system(1) from (35):
\be
\alf=2\frac{k\theta}{cosh^2(kx+k^3t)}.
\ee
To end this section, we remark that many other solutions of sKdV systems may be
obtained in this way.
\section{Conclusion}
In this paper, we constructed the supersymmetric generalizations of Darboux
transformations. Our main results are neatly presented by four propositions
given above. By studying the applications of these results, the B\"{a}cklund
transformations and some exact solutions of supersymmetric KdV systems are
obtained. Thus we that the supersymmetric versions of Darboux transformations
are indeed important.

This is just the starting point of such study and there are many questions to
be answered. For instance, it would be interesting to consider the
supersymmetric analogies of the Crum's transformations. Also, it is important
to pursue the similar results for other supersymmetric integrable models such
as supersymmetric Sine-Gordon, supersymmetric Nonlinear Schr\"{o}rdinger
equation, and supersymmetric KP hierarchy, etc.. The results along this line
may be presented elsewhere.
\par
\vspace{10pt}
{\bf{ACKNOWLEDGEMENT}}\par
The part of the work were presented during the ITP Workshop on the Frontiers in
Quantum Field Theory(Beijing) and the International Workshop on Nonlinear
Schr\"{o}dinger Equation(NLS'94, Moscow). I should like to thank the organizers
for their hospitality.. I should like also to thank A. Fordy, M. Manas, R.
Sasaki and A. Shabat for their interest in this work.\par
\vspace{.1in}
\bigskip

\large
REFENERCES
\bigskip
\par
\small
1 Athorne, C., and Nimmo, J.J.C., {\em Inverse Problems} {\bf{7}}, 648(1991);
{\bf 7}, 809 (1991).\par
2 Chau, L.L., Shaw, J.C. and Yen, H.C., {\em Commun. Math. Phys.} {\bf 149},
263(1992).\par
3 Darboux, G., {\em C. R. Acad. Sci. Paris} {\bf 94}, 1456(1882).\par
4 Figueroa-O'Farill, J., Mas, J., and Ramos, E., {\em Rev. Math. Phys.} {\bf
3}, 479(1991).\par
5 Gates Jr., S.J., and Nishino,  H., {\em Phys. Letts.} {\bf 299B}, 255(1993).
\par
6 Gu, C.H. et al, {\em Soliton Theory and its Applications}, Zhejiang
Publishing House of Science and \hspace{6mm}Technology(1990).\par
7 Inami, T., and Kanno, H., {\em Commun. Math. Phys.} {\bf 136}, 519(1991).\par
8 Kupershmidt, B.A., {\em Phys.Letts.} {\bf 102A}, 213(1984).\par
9 Levi, D., {\em Inverse Problems} {\bf 4}, 165(1988).\par
10 Levi, D., Ragnisco, O. and Sym, A., {\em IL Nvovo Cimento} {\bf 83B}
314(1984).  \par
11 Manin, Yu., and Radul, A., {\em Commun. Math. Phys.} {\bf 98}, 65(1985).\par
12 Mathieu, P., {\em J. Math. Phys.} {\bf 29}, 2499(1988).\par
13 Mathieu, P., {\em Phys.Letts.} {\bf 203B}, 287(1988); Bilal and Geveis, {\em
Phys. Letts.} {\bf 211B}, 95(1988).\par
14 V.B.Matveev and M.A.Salle, {\em Darboux Transformations and Solitons},
Springer-Verlag, (1990).\par
15 Matveev, V.P., {\em Lett. Math. Phys.} {\bf 3}, 213(1979); {\bf 3},
217(1979)(1979).\par
16 Mcarthur, I.N. and  Yung, C.M., {\em Mod. Phys. Letts.} {\bf
A8},1739(1993).\par
17 Morris, C. and Pizzocchero, L., {\em Commun. Math. Phys.} {\bf 158},
267(1993).\par
18 Oevel, W. and Popowicz, Z., {\em Commun. Math. Phys.} {\bf 139}, 441(1991).
\par
19 Oevel, W. and Rogers, C., {\em Rev. Math. Phys.} (1993).\par
20 Wadati, M., Sanuki, H. and Konno, K., {\em Prog. Theor. Phys.} {\bf 53},
419(1975). \par
21 Yamanaka, I. and Sasaki, R., {\em Prog. Theor. Phys.} {\bf 79},
1167(1988).\par
 \par
\end{document}